\documentstyle[preprint,aps]{revtex}
\begin{document}
\draft
\tightenlines

\title{Distribution Function of the End-to-End Distance
of Semiflexible Polymers}

\author{J. K. Bhattacharjee and D. Thirumalai}
\address{Institute for Physical Science and Technology \\ 
and Department of Chemistry, \\
University of Maryland, College Park, MD 20742, USA}
\author{J. D. Bryngelson}
\address{Physical~Sciences~Laboratory, \\
Division~of~Computer~Research~and~Technology\\
National~Institutes~of~Health, Bethesda, MD 20892, USA}

\maketitle

\begin{abstract}
The distribution function of the end-to-end distance of a semiflexible polymer,
$G(R;L)$ (where $R$ denotes the end-to-end distance and $L$ the contour
length), is calculated using a meanfield-like approach.  The theory yields 
an extremely simple expression for $G(R;L)$ which is in excellent 
agreement with Monte Carlo simulations.  The second and fourth moments of
$G(R;L)$ agree with exact results for a semiflexible polymer in both the random
coil and the rod limit.
\end{abstract}

\pacs{   }

Many properties of isolated, flexible polymer molecules are now 
understood \cite{ref_1}.  For example, theoretical methods can be used to 
calculate the distribution function of flexible polymer molecules to very
high accuracy.  Unfortunately, many polymer molecules have too much internal
stiffness to be successfully modeled as flexible chains \cite{ref_2}.  This is 
especially true for several important biopolymers such as actin, DNA, and
microtubules \cite{ref_3}.  A measure of the stiffness of a polymer is the
persistence length, $l_p$.  To understand the persistence length, consider
two points on the polymer separated by a length $l$ along the
contour of the polymer backbone and construct the tangent vectors to this 
contour.  The persistence length is the length along the contour over which
these tangent vectors become uncorrelated. 
Thus, for $l<l_p$ the tangent vectors would
have significant correlation 
and for $l>l_p$ the tangent vectors would have little
correlation.  Typical values of $l_p$ for biopolymers range from several nm
to a few mm.  If the contour length $L$ of a polymer is of the same order
of magnitude as $l_p$ (or smaller) then the flexible chain model is inadequate
for describing the polymer.  For such a polymer it is imperative to include 
bending rigidity to describe the conformations of the chain.  An appropriate
model, called the semiflexible or wormlike model, was introduced in 1949
by Kratky and Porod \cite{ref_4}.  
The semiflexible model has been shown to provide a good 
starting point in the description of polymers with significant internal 
stiffness.

In contrast to the situation of isolated, flexible polymers, there are many
properties of semiflexible polymers that are not understood.  Inspired by 
recent experiments \cite{ref_5,ref_6} that have probed the properties of 
semiflexible biological molecules, there has been a renewed interest in 
understanding their shapes.  A central quantity for understanding the shape
of a polymer is the distribution of end-to-end distances.  Such a distribution
function can, in principle, be measured by scattering experiments.  
Furthermore, it can be used to calculate the structure factor, which is needed
as input in describing the dynamics of polymers.  Therefore, finding a 
simple expression
for this distribution function which is also accurate over a wide
range of stiffness is important for further theoretical progress in
understanding the properties of stiff chains.  This paper provides just such
a simple expression for the distribution of end-to-end distance of an ideal,
semiflexible polymer chain.

There have been several studies of the distribution of the end-to-end
distance of semiflexible chains \cite{ref_7,ref_8,ref_9}.  
Most of these studies have started by considering the chains near the rod 
limit and have computed corrections in powers of $t^{-1}$, where $t$ 
represents the ratio of the contour length to the persistence length.
These calculations are very complicated and, more importantly, they do not
provide reliable results in the interesting case where $t$ 
is of the order one.  Recently \cite{ref_10}, Wilhelm and Frey [WF] have 
reported careful analytic and numerical (Monte Carlo) calculations for the 
radial distribution function of the end-to-end distance for a range of values
of $t$.  Their analytic expressions consist of infinite series of 
parabolic cylinder functions (for two dimensions) and Hermite polynomials
(for three dimensions).  These expressions compare extremely well with 
their simulation results.  Our approach differs from these previous studies.
In this paper we use the method of our recent meanfield theory of semiflexible
polymers \cite{ref_11} to obtain a simple expression for the distribution
function of the end-to-end distance.  This expression is surprisingly
accurate when compared to the analytic theory and simulations results for
the range of $t$ studied in \cite{ref_10}.

The Hamiltonian for the semiflexible chain is taken to be \cite{ref_12}
\begin{equation}
{\cal H} = \frac{l_p}{2} \int_{0}^{L} ds \left[ 
\frac{\partial {\bf u}(s)}{\partial s} \right]^2
\label{hamiltonian}
\end{equation}
where $l_p$ ($=\kappa/k_B T$ with $\kappa$ being the bending rigidity) is the
persistence length, ${\bf u}(s)$ ($= \partial {\bf r}(s)/\partial s$) is the 
unit tangent vector to the curve ${\bf r}(s)$ which describes the chain 
contour, $s$ is the position measured
along the chain contour, and $L$ is the contour
length of the entire chain.  The distribution function of the end-to-end
vector ${\bf R}$ is 
\begin{equation}
G({\bf R};L) = \left \langle \delta \left({\bf R} - \int_0^L {\bf u}(s) ds 
	       \right) \right \rangle
\label{def_of_G}
\end{equation}
where the average is evaluated with respect to a thermal weight function, 
$\Psi[{\bf u}(s)]$,
\begin{equation}
\left \langle \ldots \right \rangle =
\frac{\int {\cal D}[{\bf u}(s)] \ldots \Psi[{\bf u}(s)]}
     {\int {\cal D}[{\bf u}(s)] \Psi[{\bf u}(s)]}
\label{def_of_average} 
\end{equation}
The thermal weight function for the semiflexible chain is
\begin{equation}
\Psi[{\bf u}(s)] = \delta({\bf u}(s)^2 - 1) \exp(-{\cal H}[{\bf u}(s)])
\label{stiff_chain_weight}
\end{equation}

In an earlier paper \cite{ref_11} we showed that when the weight
$\Psi[{\bf u}(s)]$ is replaced by 
\begin{equation}
\Psi_{MF}[{\bf u}(s)] = \exp \left\{ 
- \frac{l_p}{2} \int_0^L \left(\frac{d{\bf u}}{ds}\right)^2 ds
- \lambda \int_0^L [{\bf u}(s)^2 - 1] ds 
- \epsilon [{\bf u}(0)^2 - 1] - \epsilon [{\bf u}(L)^2 - 1]
\right\}
\label{MF_weight}
\end{equation}
and the parameters $\lambda$ and $\epsilon$ are chosen variationally 
through a stationary phase approximation, 
then the hard constraint ${\bf u}(s)^2 = 1$ is replaced by the thermally
averaged constraint $ \langle {\bf u}(s)^2 \rangle = 1$.
This result had been conjectured by Lagowski {\em et al.} \cite{ref_13}.
We also established that imposing
the constraint $ \langle {\bf u}(s)^2 \rangle = 1$ for every point 
on the chain requires two parameters; specifically, the parameter $\epsilon$
is required to suppress the fluctuations at the ends of the
chain \cite{ref_11,ref_13}.  
Furthermore, if one uses only the variational parameter $\lambda$,
as suggested elsewhere \cite{vilgus}, then one obtains a Gaussian expression
for $G(R;L)$ which is wrong.
In more recent work \cite{ref_14} we showed that the same strategy, {\em i.e.},
replacing the exact thermal weight by Eq. (\ref{MF_weight}) and evaluating
$\lambda$ and $\epsilon$ variationally, produces
excellent quantitative results for the elastic response of a semiflexible
chain under tension.
Notice that the optimal values of $\lambda$ and $\epsilon$ depend on the
property of interest.

Following these earlier works \cite{ref_11,ref_14} we shall calculate 
$G(R;L)$ by replacing the true thermal weight in
Eq. (\ref{stiff_chain_weight}) by $\Psi_{MF}[{\bf u}(s)]$ in
Eq. (\ref{MF_weight}) and use a stationary phase approximation to set
$\lambda$ and $\epsilon$.
The equation for $G(R;L)$ with the weight given by $\Psi_{MF}[{\bf u}(s)]$ 
is
\begin{equation}
G(R;L) = \Omega \int_{-i \infty}^{+i \infty} \frac{d^3{\bf k}}{(2 \pi i)^3}
\int d \lambda \int d \epsilon \int {\cal D}[{\bf u}(s)] 
\exp \left\{ - {\bf k} \cdot \left[{\bf R} - \int_0^L {\bf u}(s) ds 
\right] \right\} 
\Psi_{MF}[{\bf u}(s)]
\label{G_MF_1}
\end{equation}
where $\Omega$ represents a normalization constant.
The functional integral over ${\bf u}(s)$ in Eq. (\ref{G_MF_1}) is done by
replacing ${\bf u}$ by ${\bf v} = {\bf u} - {\bf k}/2 \lambda$.  The 
resulting path integral corresponds to a harmonic oscillator that makes 
a transition from ${\bf u}(0) - {\bf k}/2 \lambda$
to ${\bf u}(L) - {\bf k}/2 \lambda$ in imaginary `` time" $L$.
Using the standard result for the harmonic oscillator propagator \cite{ref_15}
the distribution function $G(R;L)$ becomes
\begin{equation}
G(R;L) = \int d \lambda \int d \epsilon \exp(-F(\lambda,\epsilon))
\end{equation}
where
\begin{eqnarray}
F(\lambda, \epsilon) &=& \frac{3}{2} \left[ 
   \log \left( \frac{\lambda L - 2 \epsilon}{4 \lambda^2} \right)
 + \log \left( \frac{\sinh (\omega L)}{\omega l_p} \right)
 + \log ( \alpha \beta )
 \right] \nonumber \\
 &-& \lambda L - 2 \epsilon
 + \left( \frac{\gamma}{\beta} \right)
   \left( \frac{\lambda^2 R^2}{\lambda L - 2 \epsilon} \right)
 + \mbox{const.}
\label{F}
\end{eqnarray}
with
\begin{eqnarray}
\alpha &=& \frac{\epsilon}{2} 
	+  \frac{l_p \omega}{4} \coth \left( \frac{\omega L}{2} \right) \\
\beta  &=& \frac{\epsilon}{2}
	+  \frac{l_p \omega}{4} \tanh \left( \frac{\omega L}{2} \right)
	+  \frac{1}{\lambda L - 2 \epsilon}  \\
\gamma &=& \frac{\epsilon}{2}
	+  \frac{l_p \omega}{4} \tanh \left( \frac{\omega L}{2} \right)
\end{eqnarray}
and 
\begin{equation}
\omega = \left( \frac{2 \lambda}{l_p} \right)^{1/2}
\end{equation}
For the case of large $L$ the function $F(\lambda, \epsilon)$ may be written 
in the somewhat more manageable form
\begin{equation}
F(\lambda, \epsilon) = Lf(\lambda) + \frac{3}{2} \log(2 \epsilon + l_p \omega)
- 2 \epsilon - \frac{9}{2} \log(l_p \omega) 
+ {\rm O} \left( \frac{1}{L} \right) + \mbox{const.}
\label{F_approx}
\end{equation}
where $f(\lambda) = (3/2) \omega - \lambda (1-r^2)$ and $r = R/L$.
Thus, as $L$ becomes larger the function $F(\lambda, \epsilon)$ becomes more
sharply peaked and the stationary phase approximation becomes more
accurate.  For leading order in $L$ the stationarity condition for $\lambda$
is $f'(\lambda) = 0$, which gives
\begin{equation}
\left(\frac{\lambda l_p}{2}\right)^{1/2} = 
\frac{3}{4} \left(\frac{1}{1-r^2}\right)
\end{equation}
and we find that $\epsilon$ does not show any significant, leading order
variation with $r$.  Substituting the stationary values of $\lambda$ and
$\epsilon$ into Eq. (\ref{F_approx}) for $F(\lambda, \epsilon)$
yields the desired simple, approximate
expression for $G(R;L)$,
\begin{equation}
G(R;L) = \frac{N}{(1-r^2)^{9/2}} \exp \left( -\frac{9L}{8 l_p (1-r^2)} \right)
\end{equation}
where $N$ is a normalization constant and we have again taken the leading
orders in $L$ in the exponential and in the prefactor.
In our earlier work \cite{ref_11}
we showed that the stationary phase approximation 
reduces the persistence length from $l_p$, which obtains in an exact
treatment of the thermal weight in Eq. (\ref{stiff_chain_weight}),
to $l_{MF} = (2/3)l_p$.  Identifying the effective persistence length
$l_{MF}$ with the measured persistence length of the polymer produces a 
simple approximate expression for the radial probability density of a 
semiflexible chain in three dimensions,
\begin{equation}
P(r;t) = \frac{4 \pi N r^2}{(1-r^2)^{9/2}}
	 \exp\left( - \frac{3t}{4}\frac{1}{(1-r^2)} \right)
\label{radial}
\end{equation}
where $t= L/l_{MF}$.  The normalization constant is determined by the
requirement that 
\begin{equation}
\int_0^1 P(r;t) dr = 1.
\label{normal_condition}
\end{equation}
The integral can be evaluated by the substitution $r = x/\sqrt{1+x^2}$
to yield
\begin{equation}
N = \frac{4 \alpha^{3/2} e^{\alpha}}{\pi^{3/2}(4 + 12 \alpha^{-1}
	                          + 15 \alpha^{-2})}
\end{equation}
where $\alpha = 3t/4$.

The distribution function, $P(r;t)$, vanishes as $r^2$ as $r \rightarrow 0$ 
and also vanishes at $r=1$.  The peak of the distribution function occurs at
$r_{max} = [(\eta + \sqrt{\eta^2 + 14})/7]^{1/2}$, 
where $\eta = (5/2) - \alpha$.  As expected, when $t \rightarrow 0$, 
then $r_{max} \rightarrow 1$ and $P(r,t) \rightarrow \delta(r-1)$.
In Fig. (1) we plot $P(r;t)$ for the five values of $t$ for which WF
\cite{ref_10} have presented simulation data.  For comparison, the
results of the analytic expressions obtained by WF \cite{ref_10}
are also presented.  For very stiff chains ($t=0.5$ in Fig. (1)) the WF
theory gives a more accurate estimate of the peak \cite{ref_17}. 
Nevertheless, over the range of $t$ examined by WF \cite{ref_10} our simple
expression in Eq. (\ref{radial}) reproduces the data quite accurately.

In order to further assess the validity of our theory we have also 
calculated the second and fourth moments using the radial distribution
function in (\ref{radial}).  The moments are given by
\begin{equation}
\mu_n \equiv \langle r^n \rangle = \int_0^1 r^n P(r;t) dr.
\end{equation}
The same substitution that was used to evaluate the integral in 
Eq. (\ref{normal_condition}) can be used to find the moments, yielding
\begin{eqnarray}
\mu_2 &=& \left( \frac{3}{2} \right)
          \frac{4 \alpha + 10}{4 \alpha^2 + 12 \alpha +15} \\
\mu_4 &=& \frac{15}{4 \alpha^2 + 12 \alpha +15}
\end{eqnarray}
for the second and fourth moments.
In the limit of $\alpha \rightarrow \infty$ (random coil limit)
$\mu_2 \rightarrow 3/2\alpha$ and $\mu_4 \rightarrow 15/4\alpha^2$, both
of which coincide with results using exact moments \cite{ref_12}.
Similarly both $\mu_2$ and $\mu_4$ tend to unity in the limit of 
$\alpha \rightarrow 1$ (hard rod limit), which is once again the 
exact result.

The theory outlined here provides surprisingly accurate results for $P(r;t)$.
The correct limiting behavior is obtained for $P(r;t)$ and for the first 
two moments.  Furthermore, our theory can be systematically applied to 
other problems involving semiflexible chains.  Thus it appears that the 
meanfield-like approach used here may be useful for treating a wide 
variety of problems in which the rigorous enforcement of the constraint
${\bf u}(s)^2 = 1$ is difficult to enforce.

We wish to thank Dr. G. H. Weiss for useful suggestions and a referee for 
pointing out a misprint in an earlier version of this paper.

\newpage
\noindent Figure Caption \\ 

\noindent Fig. 1: Comparison of $P(r,t)$ obtained 
with Monte Carlo simulations (represented by symbols) 
for $t=10$, $5$, $2$. $1$, and $0.5$
and with analytic theories (represented by curves) for the same values of $t$.
The curves and symbols are arranged so that the largest value of $t$ is the
left most and the smallest value of $t$ is the right most.  The dark lines
corresponds to Eq. (\ref{radial}) and the light lines are based on the 
approximate theory of Wilhelm and Frey \cite{ref_10}.


\begin{references}
\bibitem{ref_17}
In the extreme limit $t \rightarrow 0$ we expect 
$P(r;t) \sim \beta (t) \delta (1-r)$ where $\beta (t)$ is the area under 
the $\delta$-function.  It is clear that Eq. (\ref{radial}) 
goes to the correct limit
as $t \rightarrow 0$.  For very small $t$, $\beta (t)$ would be approximately
the area under a rather sharply peaked function at $r=1$ (see the 
$t=0.5$ case in Fig. (1)).  Although our theory overestimates the peak by 
about ten percent it is clear that the area under $P(r;t)$ and the width are
very close to the simulation results.
\end{references}
\end{document}